\documentclass[twocolumn,floatfix, showpacs]{revtex4}

\usepackage[final]{epsfig}
\usepackage{amsmath}
\begin{document}
 
\title{The Physics probed by the $P_T$ Dependence of the Nuclear Suppression Factor}
 
\author{Thorsten Renk}
\email{thorsten.i.renk@jyu.fi}
\affiliation{Department of Physics, P.O. Box 35, FI-40014 University of Jyv\"askyl\"a, Finland}
\affiliation{Helsinki Institute of Physics, P.O. Box 64, FI-00014 University of Helsinki, Finland}


\begin{abstract}
The nuclear suppression factor $R_{AA}$ of single inclusive hadrons measured in ultrarelativistic heavy-ion collisions was the first observable to study jet quenching, i.e. the final state interaction of hard parton showers with the surrounding bulk matter. While its transverse momentum ($P_T$) dependence of $R_{AA}$ observed at RHIC was weak and hence never decisive in constraining models, there is now a strong and non-trivial $P_T$ dependence observed at the LHC. This has been a challenge for several models which worked well at RHIC kinematics. However, in more general terms it is also of importance to understand what physical properties of the hard process and the parton-medium interaction are reflected in $R_{AA}(P_T)$. The results of the work presented here suggest that the two main effects underlying the $P_T$ dependence are the Quantum-Chromodynamics scale evolution of the fragmentation function and the limited distance (set by the typical medium length scale) for which a shower evolves in the medium.

\end{abstract}
 
\maketitle

\section{Introduction}

The single inclusive hadron nuclear suppression factor $R_{AA}$ is defined as the hadron yield in heavy-ion (A-A) collisions of given centrality, divided by the yield in p-p collisions, scaled with the number of binary nucleon-nucleon collisions occurring in the heavy-ion case,
\begin{equation}
\label{E-RAA}
R_{AA}(P_T,y) = \frac{dN^h_{AA}/dP_Tdy }{T_{AA}({\bf b}) d\sigma^{pp}/dP_Tdy}.
\end{equation}
It is by construction unity if there are neither initial nor final state interactions affecting hard Quantum Chromodynamics (QCD) processes in A-A collisions. Experimentally, the indication of $R_{AA} \ll 1$ has been taken as one of the crucial verifications of the collective nature of A-A collisions, as such an observation indicates significant final state interaction of hard partons with hot and dense matter.

Results for $R_{AA}$ were the first indication of jet quenching in the RHIC era \cite{PHENIX-RAA} and among the first reported findings at the start of the LHC heavy-ion program \cite{ALICE-RAA}. Unlike at RHIC, the first results from LHC showed a strong rise of  $R_{AA}$ with $P_T$ with a minimum well below the RHIC value. Later results from the CMS collaboration out to 100 GeV \cite{CMS-RAA} then showed that the rise does not continue but rather appears to level off from about 40 GeV onward at a value of $\sim 0.6$.

The strong rise of $R_{AA}$ was unexpected for a number of models which tended to overestimate the expected suppression at the LHC (e.g. \cite{RAA-LastCall,SurprisingTransparency}) but could be reproduced in others \cite{Majumder,Wang}. The $P_T$ dependence of $R_{AA}$ has consequently been recognized as an important tool to distinguish models which yield near-identical results at RHIC kinematics \cite{RAA-Extrapolation,Constraining}.

There is, however, significantly less information available what physics drives the $P_T$ dependence of $R_{AA}$ in various models and what particular scenarios are supported or disfavoured by the data. It is the aim of this work to illustrate the rich combination of various effects leading to the observed $P_T$ dependence at the example of the in-medium shower evolution code YaJEM \cite{YaJEM1,YaJEM2,YaJEM-DE}

\section{Qualitative estimates}

\subsection{Simple considerations}

Following the discussion in \cite{Constraining}, the most general representation of the partonic final state interaction in a hot medium is the medium-modified fragmentation function (MMFF) $D_{i \rightarrow h} (z, E_i, Q_i^2 | T_1(\zeta), T_2(\zeta), \dots T_n(\zeta))$, i.e. the distribution of hadrons $h$ given a parton $i$ with initial energy $E_i$ and initial virtuality $Q_i^2$ where the hadron energy $E_h = z E$ and the parton has traversed a medium along the path $\zeta$ where $T_i(\zeta)$ are the medium transport coefficients relevant for the process. If the MMFF can be computed, its convolution with the perturbatively calculable parton spectrum gives the medium-modified hadron yield.

A frequently used simplification in the case of leading hadron observables is the so-called energy loss approximation 

\begin{equation}
D_{i \rightarrow h} (z, E, Q_i^2 | T_i(\zeta)) \approx P(\Delta E,E|T_i(\zeta)) \otimes D_{i \rightarrow h}(z, Q_i^2)
\end{equation}

where the MMFF is replaced by an energy loss probability distribution $P(\Delta E,E|T_i(\zeta))$ for the leading parton which is convoluted with the vacuum fragmentation function $D_{i \rightarrow h}(z, Q_i^2)$ which only depends on the momentum fraction $z$ and the scale at which the process takes place.

Using a very simple constant energy loss assumption, a qualitative argument why $R_{AA}$ should generically increase with $P_T$ can be made as follows \cite{KinLimit}: Parton spectra in pQCD  can be approximated by a power law as $dN/dp_T = const./p_T^n$ where $n\approx 7$ for RHIC kinematics and $n \approx 5$ at LHC. Assuming that one can approximate the effect of the medium by the mean value energy loss $\langle \Delta E \rangle$ (for realistic energy loss models, this is not a good approximation, as fluctuations around the mean turn out to be large), the energy loss corresponds to a shift in the parton spectrum prior to fragmentation (note that this does not correspond to a realistic spacetime picture of the reaction --- the uncertainty relation suggests that the scale evolution of the fragmentation function takes place in medium). 

The shift can be taken into account by the replacement $p_T \rightarrow p_T + \langle \Delta E \rangle$ in the expression for the parton spectrum. $R_{AA}(p_T)$ can then be approximated by the ratio of the parton spectra before and after energy loss as

\begin{equation}
\label{E-RAAApprox1}
R_{AA}(p_T) \approx \left(\frac{p_T}{p_T + \langle \Delta E \rangle}\right)^n = \left(1 - \frac{\langle \Delta E \rangle}{p_T + \langle \Delta E \rangle}\right)^n
\end{equation}

and it is easily seen that this expression approaches unity for $p_T \gg \langle \Delta E\rangle$ and that $n$ governs how fast this limit is aproached.

In contrast, assuming a constant fractional energy loss $\Delta E = z p_T$ leads to a suppression factor independent of $p_T$ in a power law approximation, as

\begin{equation}
\label{E-RAAApprox2}
R_{AA}(p_T) \approx \left( \frac{p_T}{p_T + z p_T}\right)^n = \left(\frac{1}{1+z} \right)^n.
\end{equation}

However, in reality the perturbative QCD (pQCD) parton spectrum is only locally approximated by a power law and the power $n$ of a local fit to the spectrum increases with $p_T$ in the kinematically accessible range. This implies that the constant energy loss scenario approaches unity more slowly for a real spectrum than for a simple power law approximation and that constant fractional energy loss leads to a \emph{decrease} of $R_{AA}$ with $P_T$ in the realistic case (see e.g. \cite{gamma-h} for an explicit calculation).

It follows that the $P_T$ dependence of $R_{AA}$ is always determined from a \emph{combination} of parton-medium interaction physics and shape of the primary parton spectrum. 

\subsection{The role of in-medium shower evolution}

As indicated above, a more realistic treatment of parton-medium interaction involves computing the MMFF. In the following, we illustrate the chief additional  mechanisms generating $P_T$ dependence of $R_{AA}$ using the example of the in-medium shower evolution Monte-Carlo (MC) code YaJEM \cite{YaJEM1,YaJEM2,YaJEM-DE}.

In the absence of a medium, YaJEM is identical to the PYSHOW algorithm \cite{PYSHOW} which evolves partons as a series of $a\rightarrow bc$ branchings in the energy fraction $z = E_b/E_a$ and the virtuality $t = \ln(Q_a^2)/\Lambda_{QCD}^2$ with $\Lambda_{QCD} = O(300)$ MeV, starting from an initial virtuality scale $Q_i$ which is given by the momentum scale of the hard process. This series of branchings terminates at a lower scale $Q_0$, at which point non-perturbative dynamics is assumed to dominate and the Lund hadronization model \cite{Lund} is used. 

In YaJEM, it is assumed that the virtuality $Q_a^2$ and energy $E_a$ of any intermediate shower parton $a$ is modified by the medium via two transport coeffients, $\hat{q}$ and $\hat{e}$ as

\begin{equation}
\label{E-Qgain}
\Delta Q_a^2 = \int_{\tau_a^0}^{\tau_a^0 + \tau_a} d\zeta \hat{q}(\zeta)
\end{equation}

and

\begin{equation}
\label{E-Drag}
\Delta E_a = \int_{\tau_a^0}^{\tau_a^0 + \tau_a} d\zeta \hat{e}(\zeta)
\end{equation}

as the parton propagates along the path $\zeta$.
To evaluate these equations requires a mapping of the shower evolution of PYSHOW in momentum space to the hydrodynamical evolution in position space and a model of the transport coefficients as a function of thermodynamical properties of the medium. 

The temporal structure of the shower evolution can be parametrically recovered by uncertainty arguments. The mean lifetime of a virtual parton $b$  coming from a parent $a$ is hence given as 

\begin{equation}
\label{E-Lifetime}
\langle \tau_b \rangle = \frac{E_b}{Q_b^2} - \frac{E_b}{Q_a^2}.
\end{equation} 

In the MC simulation of the shower, the actual lifetime is determined from this mean value according to the probability distribution

\begin{equation}
\label{E-RLifetime}
P(\tau_b) = \exp\left[- \frac{\tau_b}{\langle \tau_b \rangle}  \right].
\end{equation}

Two important observations follow: Medium-induced changes of the shower kinematics are not significant if $Q^2 \gg \Delta Q^2$ (or $E \gg \Delta E$). Probing $R_{AA}$ at  higher $P_T$ corresponds to probing higher values of the initial virtuality scale $Q_i$ and to higher values of parton energies $E_i$, and thus both the partonic shower evolution as well as the part of the shower evolution in which $Q^2 \gg \Delta Q^2$ grow with $P_T$ (this is the MC shower equivalent of the QCD scale evolution of the fragmentation function). As a result, vacuum and on-medium shower evolution become progressively similar at high $P_T$ with the implication $R_{AA} \rightarrow 1$ (see \cite{KinLimit} for an explicit computation) --- the QCD scale evolution thus has a generic tendency to increase $R_{AA}$.

At the same time, effects at the lower scale $Q_0$ also influence $R_{AA}$. As Eq.~(\ref{E-Lifetime}) indicates, at sufficiently high $E$ the formation length of a branching eventually exceeds the dimenstions of the medium. If the medium is characterized by a length scale $L$, the in-medium evolution of the shower will thus proceed only down to a scale $Q_0 = \sqrt{E/L}$ \cite{Abhijit1,Abhijit2}. This in turn implies that for jets with large $E$ the intermediate virtuality $Q$ will never be small while the jet is still in medium and hence $Q^2 \gg \Delta Q^2$ will always be realized. In essence, the fact that the length scale for shower evolution can exceed the medium dimensions also implies $R_{AA} \rightarrow 1$ for sufficiently large $P_T$ \cite{YaJEM-DE}.

While these two mechanisms have been discussed here in the specific context of YaJEM, they are in fact rather generic. The lower scale being set by the medium dimensions is a consequence of the uncertainty principle and thus difficult to avoid in any model, while the weakening influence of the medium for longer QCD scale evolution is based on a scale comparison and closely analoguous to the loss of sensitvity to the low $Q^2$ non-perturbative parton distributions when an inital scale evolution is carried out to a high virtuality (see for instance \cite{EPS09}).

\subsection{Minor effects}

There are yet more effects which shape the $P_T$ dependence of $R_{AA}$ to some degree. One of them is the composition of the primary parton spectrum. Since gluons carry a color charge different from quarks and interact with a factor $C_F = 9/4$ more strongly, gluons also experience stronger medium-induced suppression than quarks. However, the ratio of quarks to gluons in the pQCD parton spectrum varies as a function of $p_T$, while the low $p_T$ part is dominated by gluon production, at high $p_T$ quark production takes over. Corresponding to this transition, there is a slight rise in $R_{AA}$ reflecting the lower interaction strength of quarks \cite{KinLimit}. 

Similarly, while one usually attributes the modification of inclusive hard hadron spectra to the final state interaction with the hot and dense medium which is not present in the case of p-p collisions, there is also a modest change of the initial state comparing p-p with A-A collisions, which is parametrized by the nuclear parton distribution functions (nPDFs), e.g. \cite{EPS09,NPDF,EKS98}. The $x$-dependence of the nPDFs, mapped to a $p_T$ dependence in the parton spectrum after convolution of the initial state parton distributions with the perturbative hard scattering cross section, leads to a non-trivial $P_T$ dependence of $R_{AA}$ even in the absence of final state interactions and e.g. a sizable enhancement of A-A over p-p near the kinematic limit \cite{KinLimit}. 

In the kinematic range of current LHC measurements of $R_{AA}$ however, both the transition of the dominant parton type and the nuclear initial state effects change $R_{AA}$ by less than 5\%, hence we will in the following take the effects consistently  into account  but not discuss them in further detail.

\section{Detailed Modelling}

\subsection{Embedding into fluid dynamics}

In order to obtain the medium-modified hadron yield, the MMFF must be averaged over the medium geometry. In this work, the medium evolution is taken to be a constrained 2+1D hydrodynamics evolution, extrapolated from RHIC to LHC kinematics using the EKRT model \cite{RAA-Extrapolation}. It is of some importance that a realistically evolving medium is utilized as a background, since embedding into e.g. a static background can lead to a core-corona scenario \cite{h-h} in which partons inside a certain geometrical region are strongly suppressed whereas partons produced in the corona escape the medium unmodified. Such a geometrical suppression picture can alter the $P_T$ dependence of $R_{AA}$ \cite{gamma-h}, however this is an unphysical artefact of unrealistic modelling.

The probability density $P(x_0, y_0)$ for finding a hard vertex at the transverse position ${\bf r_0} = (x_0,y_0)$ and impact parameter ${\bf b}$ is in leading order pQCD given by the product of the nuclear profile functions as

\begin{equation}
\label{E-Profile}
P(x_0,y_0) = \frac{T_{A}({\bf r_0 + b/2}) T_A(\bf r_0 - b/2)}{T_{AA}({\bf b})},
\end{equation}

where the thickness function is given in terms of Woods-Saxon the nuclear density
$\rho_{A}({\bf r},z)$ as $T_{A}({\bf r})=\int dz \rho_{A}({\bf r},z)$ and $T_{AA}({\bf b})$ is the standard nuclear overlap function $T_{AA}({\bf b}) = d^2 {\bf s}\, T_A({\bf s}) T_A({\bf s}-{\bf b})$. 

If the angle between outgoing parton and the reaction plane is $\phi$, the path of a given parton through the medium $\zeta(\tau)$, i.e. its trajectory $\zeta$ as a function of proper medium evolution time $\tau$ is determined in an eikonal approximation by its initial position ${\bf r_0}$  and the angle $\phi$ as $\zeta(\tau) = \left(x_0 + \tau \cos(\phi), y_0 + \tau \sin(\phi)\right)$ where the parton is assumed to move with the speed of light $c=1$ and the $x$-direction is chosen to be in the reaction plane. 

The transport coefficients used in YaJEM are then obtained from the hydrodynamical energy density $\epsilon$ as 

\begin{equation}
\label{E-qhat}
\hat{q}|E(\zeta) = K|K_E \cdot 2 \cdot \epsilon^{3/4}(\zeta) (\cosh \rho - \sinh \rho \cos\alpha)
\end{equation}

where $K$ and $K_E = 0.08 K$ are parameters determining the overall normalization of the coefficients, $\rho$ is the local flow velocity of the medium and $\alpha$ is the angle between parton propagation direction and flow direction. In the following, in order to focus on the $P_T$ dependence of $R_{AA}$ and make the various influences discussed below comparable, $K$ is always adjusted such that $R_{AA}$ at 10 GeV reproduces the data.

If $D_{i\rightarrow h}(z, E_i, Q_i|\zeta)$ is the MMFF as computed in YaJEM for the path $\zeta$, then the medium-averaged MMFF is obtained as

\begin{equation}
\label{E-D_TAA}
\begin{split}
\langle D_{i\rightarrow h}&(z,E_i,Q_i)\rangle_{T_{AA}} \negthickspace =\\ &\negthickspace \frac{1}{2\pi} \int_0^{2\pi}  
\negthickspace \negthickspace \negthickspace d\phi 
\int_{-\infty}^{\infty} \negthickspace \negthickspace \negthickspace \negthickspace dx_0 
\int_{-\infty}^{\infty} \negthickspace \negthickspace \negthickspace \negthickspace dy_0 P(x_0,y_0)  
D_{i\rightarrow h}(z, E_i, Q_i|\zeta)
\end{split}
\end{equation}

From this, the medium-modified production of hadrons is obtained from

\begin{equation}
\label{E-Conv}
d\sigma_{med}^{AA\rightarrow h+X} \negthickspace \negthickspace = \sum_i d\sigma_{vac}^{AA \rightarrow i +X} \otimes \langle D_{i\rightarrow h}(z,E_i,Q_i)\rangle_{T_{AA}}
\end{equation} 

from where $R_{AA}$ can be obtained.

\subsection{The role of the evolution scales}

As discussed above, the MMFF is obtained from a partonic shower evolution in medium starting from a virtuality scale $Q_i$ and extending down to a scale $Q_0 = \sqrt{E/L}$, followed by a further evolution in vacuum down to the non-perturbative hadronization scale $Q_h$, taken to be 1 GeV in the following. In order to study the importance of the evolution of these scales with $P_T$, let us first consider the outcome of a calculation where the scales are held fixed.

\begin{figure}[htb]
\epsfig{file=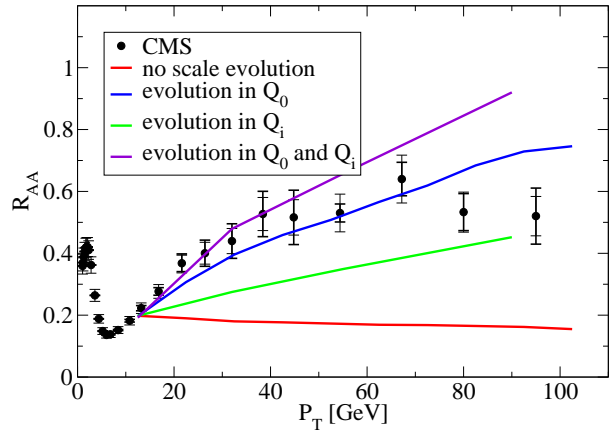, width=8cm}
\caption{\label{F-1}$P_T$ dependence of $R_{AA}$ as obtained by the CMS collaboration \cite{CMS-RAA} in comparison with YaJEM results with different treatment of upper and lower partonic evolution scale.}
\end{figure}

Using $Q_i = 20$ GeV, $Q_0 = Q_h = 1$ GeV, we generate a MMFF without scale dependence which is used throughout the kinematic range considered. Since this resulting MMFF, once averaged over the medium, retains only a dependence on the fractional momentum $z$, all medium modification is necessarily formulated as function of $z$ as well. The alert reader will realize that this corresponds closely to a constant fractional energy loss scenario in which a \emph{decrease} of $R_{AA}$ is expected after folding with a pQCD parton spectrum. As Fig.~\ref{F-1} shows, this is indeed the case, resulting in a curve in striking disagreement with the data (note that the earliest YaJEM results did not include scale evolution and show this trend \cite{YaJEM1}). 

If the in-medium evolution is not terminated at $Q_0 = 1$ GeV but rather at the scale $Q_0 = \sqrt{E/L}$ as dynamically computed given the parton initial position, path and energy in the MC simulation, with the initial scale $Q_i$ still held fixed at 30 GeV and vacuum evolution carried out between $Q_0 $ and $Q_h$, the trend reverts to a rise with $P_T$ and reasonable agreement with the data is found.

Likewise, if the lower scale is fixed to $Q_0 = 1$ GeV but $Q_i$ is allowed to vary with initial parton energy $E$ (corresponding to the usual QCD scale evolution of the fragmentation function), the decreasing trend reverts to a rising trend, albeit a weaker one as in the previous case which is not in agreement with the data.

Finally, in the realistic case in which both scales are allowed to vary, a strong rise of $R_{AA}$ with $P_T$ is observed, overshooting  the data in the region beyond 40 GeV (note that this corresponds to the curve published in \cite{RAA-Extrapolation}.

While these results establish that the treatment of the shower evolution scales is a crucial influence on the $P_T$ dependence of $R_{AA}$, the outcome that the most realistic evolution scenario overshoots the data is somewhat unsatisfactory.

\subsection{A closer look at the lower evolution scale}

While the QCD evolution of the MMFF in terms of $Q_i$ is a concept which is well understood in vacuum QCD, the determination of the lower scale $Q_0$ is not on the same level or rigor. The expression $Q_0 = \sqrt{E/L}$ is parametrically set by the uncertainty relation, but this implies that the expression is valid up to a factor $O(1)$ and hence begs the question of precisely what $E$ and what $L$ should be inserted into it.

For instance, rather than estimating the lower scale for the whole jet (using the shower-initiating parton energy $E$) we might also the leading subjet based on the argument that we are considering an observable which is predominantly sensitive to the fragmentation of leading partons, and thus what determines the medium-modification of these is only the subjet in which the leading parton is found. This would argue for a relation $Q_0 = \sqrt{f E/L}$ with $f \approx 0.7$ at RHIC and $f\approx 0.5$ at LHC.

Likewise, the in-medium distance is typically taken as the distance of the hard reaction vertex to the Cooper-Frye surface of the hydrodynamically evolving medium. However, as discussed in \cite{HydSys}, there is no compelling reason to use precisely the Cooper-Frye surface. In the context of fluid dynamics, the Cooper-Frye surface is usually taken to be an isothermal surface representing an idealized separation of matter which is interacting so strongly that fluid dynamics is justified and matter interacting so weakly that free streaming is valid. In contrast, jet-medium interactions do not require medium particles to interact, they merely require the presence of scattering centers, and hard partons would interact also with a free streaming hadronic medium. This may indicate that a realistic $L$ would be somewhat larger than the distance to the Cooper-Frye surface.

\begin{figure}[htb]
\epsfig{file=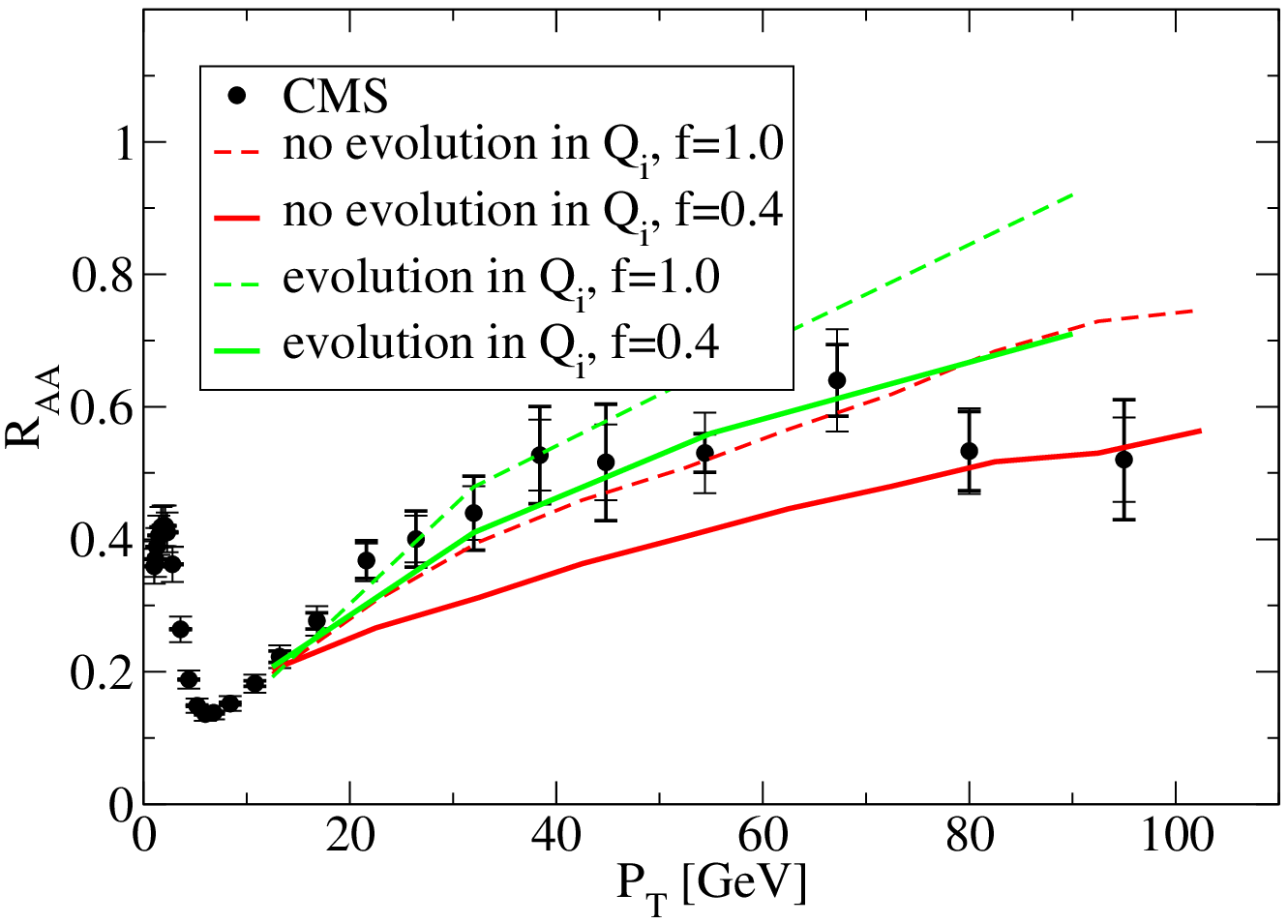, width=8cm}
\caption{\label{F-2}$P_T$ dependence of $R_{AA}$ as obtained by the CMS collaboration \cite{CMS-RAA} in comparison with YaJEM results varying the parametric estimate for the lower scale.}
\end{figure}

Estimating the decoupling of the jet to be on average 20\% beyond the Cooper-Frye surface and taking only the subjet of the leading parton into account results in a plausible scenario with $f= 0.4$ which we explore in Fig.~\ref{F-2}.

It is apparent from the figure that using $f=0.4$ is just enough to compensate the effect of the QCD scale evolution in the kinematic range considered, resulting in a scenario which is in agreement with the data up to 70 GeV while still overshooting the last two data points. Thus, within plausible variations of the parametric expression for the lower scale, the realistic computation gives a fair account of the data.

\subsection{The plateau region}

The question remains whether the trend indicated by the data is a plateau at high $P_T$ or not. There is no easy theoretical argument why $R_{AA}$ should reach a plateau at around 40-60 GeV. Both upper and lower scale evolution arguments asymptotically imply $R_{AA} \rightarrow 1$. The increasing power $n(p_T)$ of local power law fits to the pQCD spectrum imply also a decreasing sensitivity to parton energy loss at least up to $\sqrt{s}/4$ (beyond phase space restrictions become important \cite{KinLimit}). Neither a constant energy loss nor a constant fractional energy loss would thus result in a constant $R_{AA}$, rather achieving a constant $R_{AA}$ would require fine-tuning of the parton-medium interaction strength to just compensate for the effects from the shape of the parton spectrum and the scale evolutions.

One idea which could provide such a mechanism is the energy dependence of the transport coefficient $\hat{q}$ as suggested in \cite{qhat-edep}. In order to explore this idea, we parametrize the result of $\hat{q}(E)$ for a temperature of 400 MeV from Fig.~9 in \cite{qhat-edep} as
 
\begin{equation}
\hat{q}(E) = \hat{q} (10 \quad \text{GeV}) \cdot \left(1.8  \frac{E}{1 \,\text{GeV}} - 1.8\right)
\end{equation}

\begin{figure}[htb]
\epsfig{file=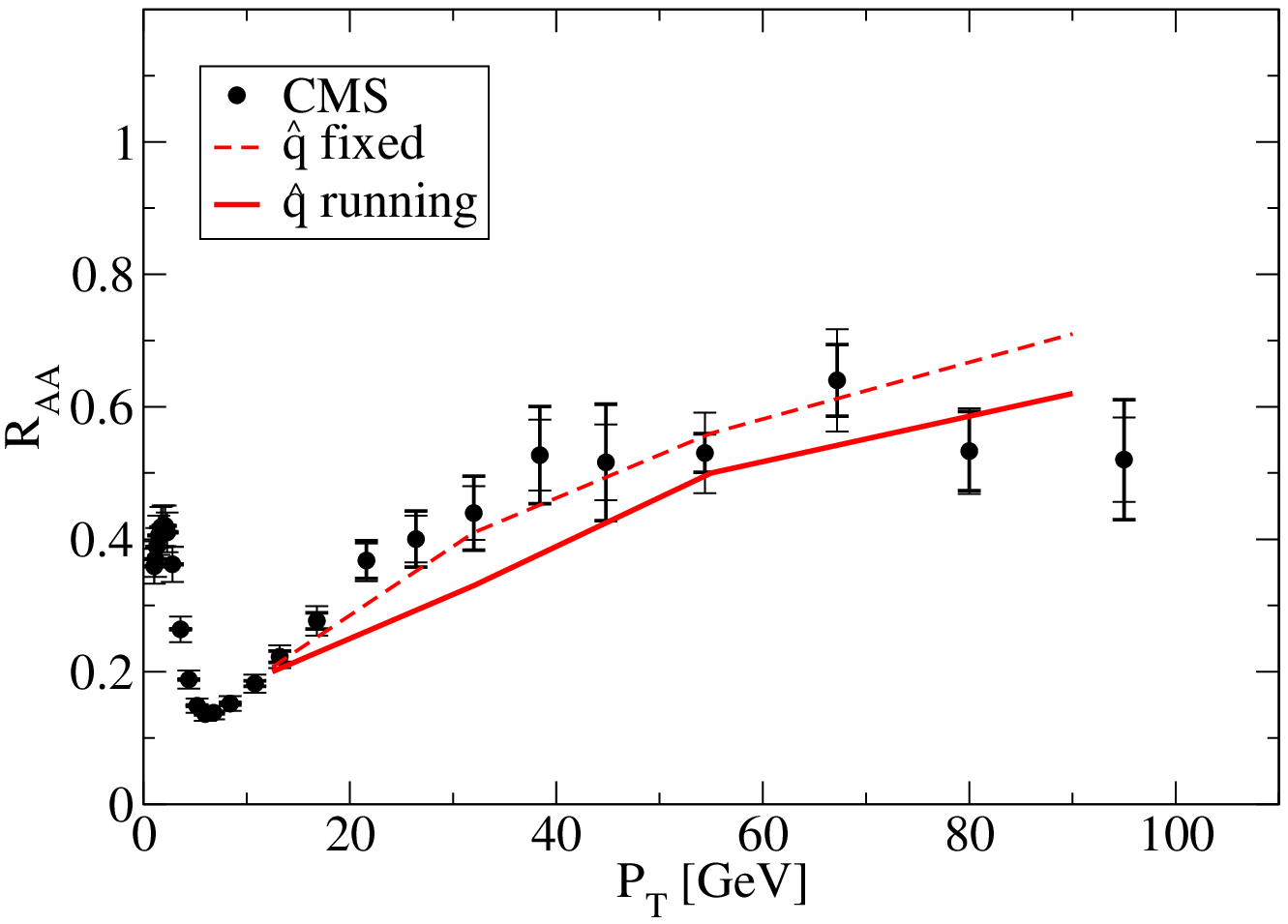, width=8cm}
\caption{\label{F-3}$P_T$ dependence of $R_{AA}$ as obtained by the CMS collaboration \cite{CMS-RAA} in comparison with YaJEM results varying the parametric estimate for the lower scale.}
\end{figure}

(note that this neglects the temperature dependence of the evolution of the transport coefficient) and use this expression to compute again hadron $R_{AA}$. The result is shown in Fig.~\ref{F-3}.

The factor $\sim$ 2 increase in $\hat{q}$ over the kinematic range leads to better agreement with the data points in the 'plateau' region above 80 GeV, however the overall shape agreement between data and calculation worsens (note again that both curves are normalized to the data at $P_T = 10$ GeV). This indicates again that generating a plateau in single inclusive $R_{AA}$ is in essence a fine-tuning problem.

\section{Considering jet $R_{AA}$}

The CMS collaboration has also measured the nuclear suppression factor of jets (defined via clustering with anti-$k_T$ with a radius parameter of $R=0.3$) and observed a result rather independent of $P_T$ in a kinematic region comparable to the parton kinematics underlying the hadronic $R_{AA}$. This has been taken as independent confirmation of the existence of a plateau \cite{Roland}.

One has to realize  however that the physics underlying the suppression of single inclusive hadrons is rather different from the suppression of clustered jets. Fundamentally, jets are much more robust against medium-induced radiation (or indeed any radiation process) since close-to-collinear radiation is always clustered back into the jet and only large-angle radiation processes outside the cone radius lead to a suppression of the jet rate \cite{Dijets}. There is thus no \emph{a priori} reason to expect that the mechanisms discussed above affect jet $R_{AA}$ in the same way.

\subsection{Modelling}

Jet $R_{AA}$ is obtained using the same medium evolution scenario and the same pQCD calculation for the hard process itself as described above and used to compute hadronic $R_{AA}$. In the following, the parameter $K$ describing the overall medium quenching strength (see Eq.~(\ref{E-qhat}) is taken directly from the corresponding hadron suppression scenario and no additional fit procedure for jet $R_{AA}$ is used.

The only difference is that in the following the output of YaJEM is clustered using the anti-$k_T$ algorithm of the FastJet package to obtain the probability distribution $P(z,E_i, Q_i^2| \hat{q}(\zeta), \hat{e}(\zeta) )$ to cluster the shower from a parton with initial energy $E_i$ and virtuality $Q_i$ into a jet with energy $E_{jet} = z E_i$ rather than the MMFF. This probability distribution can then be averaged over the geometry and convoluted with the spectrum in the same way as the MMFF.

\subsection{Results}

\begin{figure}[htb]
\epsfig{file=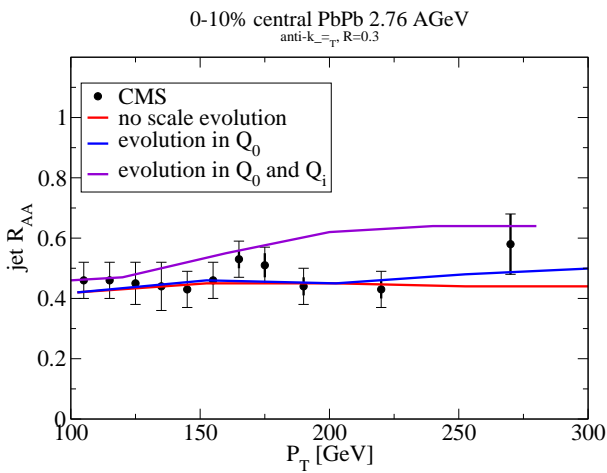, width=8cm}
\caption{\label{F-4}$P_T$ dependence of jet $R_{AA}$ as obtained by the CMS collaboration \cite{Roland} in comparison with YaJEM results with different treatment of upper and lower partonic evolution scale.}
\end{figure}

The different scenarios with upper and lower evolution scale held fixed or determined by the kinematics and spacetime position in the MC event are plotted in Fig.~\ref{F-4} in comparison with preliminary CMS data \cite{Roland}.

A striking difference to Fig.~\ref{F-1} is that jet $R_{AA}$ is almost insensitive to the lower evolution scale. The very purpose of a clustering procedure is however to suppress physics at soft, non-perturbative scales and be sensitive only to hard, perturbative physics, thus such an insensitivity is to be expected. The remaining $P_T$ dependence is chiefly driven by the pQCD scale evolution, i.e. the initial virtuality scale $Q_i$ as well as the fact that with increased jet energy jets become kinematically more and more collimated (cf. \cite{Dijet-Edep} for a discussion of the role of collimation for the dijet imbalance), implying that it becomes increasingly difficult to radiate energy out of the jet cone and leading again to a trend $R_{AA} \rightarrow 1$ asymptotically. The growth is somewhat slower than in the hadron $R_{AA}$ case.

Another crucial observation is that a plateau-like shape with a very weak $P_T$ dependence over the kinematic range considered arises naturally for jet $R_{AA}$, thus unlike in the hadron $R_{AA}$ case there is no fine-tuning problem, and thus one can not take the lack of strong $P_T$ dependence in jet $R_{AA}$ as an indication that hadronic $R_{AA}$ would level to a plateau.

No attempt to do a simultaneous global fit of hadron and jet $R_{AA}$ to the data has been made, but it is clear from the results presented here that such a fit would result in a reasonably good agreement between model and data even without novel physics like an additional energy evolution of $\hat{q}$.

\section{Conclusions}

The $P_T$ dependence of the nuclear suppression factor $R_{AA}$ is a consequence of a non-trivial interplay of several effects, chief among them the curvature of the primary pQCD spectrum, the scale QCD scale evolution of the MMFF and the fact that the in-medium evolution of showers is restricted to a length scale set by the medium size. Subleading effects include the change in dominant parton type with $P_T$ as well as effects from the nuclear initial state. Many of these effects are fairly generic and do not depend on specific assumptions about the nature of parton-medium interaction (but approximation schemes such as the energy loss approximation dropping the virtuality evolution of the shower in medium have been made in practice).

The measured $P_T$ dependence poses significant and non-trivial constraints for models, and in particular identify approximations which can not be safely made (for instance, dropping the QCD evolution of the MMFF is clearly not a justified approximation).

Within a reasonable parameter space of the precise details of embedding the shower evolution into a hydrodynamically evolving medium, fair agreement of the most realistic modelling case with both the measured single inclusive hadron and jet $R_{AA}$ can be obtained. However, if one interprets the data as exhibiting a $P_T$-independent plateau of suppression beyond 40 GeV, then conventional pQCD mechanisms such as the scale evolution and the finite length effect are insufficient to account for the data, and novel physics is required. Even in this case, $P_T$ independence requires the cancellation of the known $P_T$ dependence of QCD effects and results in essence in a fine-tuning problem. Jet $R_{AA}$ is expected to have a generically weaker $P_T$ dependence and is hence no strong indicator for the presence or absence of a plateau. In this context, it should be kept in mind that no measured p-p baseline at 2,76 GeV has been available at high $P_T$ for the CMS publication \cite{CMS-RAA}.

Future measurements with higher statistics and higher reach in $P_T$ will reveal if $R_{AA}$ can be completely described by known pQCD effects or if novel physics mechanisms are indicated by the data.

\begin{acknowledgments}
 
 This work is supported by the Academy researcher program of the
Academy of Finland, Project No. 130472. 
 
\end{acknowledgments}

\end{document}